\documentstyle[twoside]{article}

\catcode`\@=11
\long\def\@makefntext#1{
\protect\noindent \hbox to 3.2pt {\hskip-.9pt
$^{{\eightrm\@thefnmark}}$\hfil}#1\hfill}               %CAN BE USED

\def\@makefnmark{\hbox to 0pt{$^{\@thefnmark}$\hss}}    %ORIGINAL

\def\ps@myheadings{\let\@mkboth\@gobbletwo
\def\@oddhead{\hbox{}
\rightmark\hfil\eightrm\thepage}
\def\@oddfoot{}\def\@evenhead{\eightrm\thepage\hfil
\leftmark\hbox{}}\def\@evenfoot{}
\def\sectionmark##1{}\def\subsectionmark##1{}}

\oddsidemargin=\evensidemargin
\newcounter{sectionc}\newcounter{subsectionc}\newcounter{subsubsectionc}
\renewcommand{\section}[1] {\vspace{12pt}\addtocounter{sectionc}{1}
\setcounter{subsectionc}{0}\setcounter{subsubsectionc}{0}\noindent
        {\tenbf\thesectionc. #1}\par\vspace{5pt}}
\renewcommand{\subsection}[1] {\vspace{12pt}\addtocounter{subsectionc}{1}
        \setcounter{subsubsectionc}{0}\noindent
        {\bf\thesectionc.\thesubsectionc. {\kern1pt \bfit #1}}\par\vspace{5pt}}
\renewcommand{\subsubsection}[1] {\vspace{12pt}\addtocounter{subsubsectionc}{1}
        \noindent{\tenrm\thesectionc.\thesubsectionc.\thesubsubsectionc.
        {\kern1pt \tenit #1}}\par\vspace{5pt}}
\newcommand{\nonumsection}[1] {\vspace{12pt}\noindent{\tenbf #1}
        \par\vspace{5pt}}

\newcounter{appendixc}
\newcounter{subappendixc}[appendixc]
\newcounter{subsubappendixc}[subappendixc]
\renewcommand{\thesubappendixc}{\Alph{appendixc}.\arabic{subappendixc}}
\renewcommand{\thesubsubappendixc}
        {\Alph{appendixc}.\arabic{subappendixc}.\arabic{subsubappendixc}}

\renewcommand{\appendix}[1] {\vspace{12pt}
        \refstepcounter{appendixc}
        \setcounter{figure}{0}
        \setcounter{table}{0}
        \setcounter{lemma}{0}
        \setcounter{theorem}{0}
        \setcounter{corollary}{0}
        \setcounter{definition}{0}
        \setcounter{equation}{0}
        \renewcommand{\thefigure}{\Alph{appendixc}.\arabic{figure}}
        \renewcommand{\thetable}{\Alph{appendixc}.\arabic{table}}
        \renewcommand{\theappendixc}{\Alph{appendixc}}
        \renewcommand{\thelemma}{\Alph{appendixc}.\arabic{lemma}}
        \renewcommand{\thetheorem}{\Alph{appendixc}.\arabic{theorem}}
        \renewcommand{\thedefinition}{\Alph{appendixc}.\arabic{definition}}
        \renewcommand{\thecorollary}{\Alph{appendixc}.\arabic{corollary}}
        \renewcommand{\theequation}{\Alph{appendixc}.\arabic{equation}}
        \noindent{\tenbf Appendix \theappendixc #1}\par\vspace{5pt}}
\newcommand{\subappendix}[1] {\vspace{12pt}
        \refstepcounter{subappendixc}
        \noindent{\bf Appendix \thesubappendixc. {\kern1pt \bfit #1}}
        \par\vspace{5pt}}
\newcommand{\subsubappendix}[1] {\vspace{12pt}
        \refstepcounter{subsubappendixc}
        \noindent{\rm Appendix \thesubsubappendixc. {\kern1pt \tenit #1}}
        \par\vspace{5pt}}

\newcommand{\textlineskip}{\baselineskip=13pt}
\newcommand{\smalllineskip}{\baselineskip=10pt}

\def\eightcirc{
\begin{picture}(0,0)
\put(4.4,1.8){\circle{6.5}}
\end{picture}}
\def\eightcopyright{\eightcirc\kern2.7pt\hbox{\eightrm c}}

\newcommand{\copyrightheading}[1]
        {\vspace*{-2.5cm}\smalllineskip{\flushleft
        {\footnotesize International Journal of Modern Physics B, #1}\\
        {\footnotesize $\eightcopyright$\, World Scientific Publishing
         Company}\\
         }}

\def\abstracts#1#2#3{{
        \centering{\begin{minipage}{4.5in}\baselineskip=10pt\footnotesize
        \parindent=0pt #1\par
        \parindent=15pt #2\par
        \parindent=15pt #3
        \end{minipage}}\par}}

\renewenvironment{thebibliography}[1]                   %ALL CHANGES DD 13/3/92
        {\frenchspacing
         \ninerm\baselineskip=11pt
         \begin{list}{\arabic{enumi}.}
        {\usecounter{enumi}\setlength{\parsep}{0pt}
         \setlength{\leftmargin 12.7pt}{\rightmargin 0pt} %FOR 1--9 ITEMS
         \setlength{\itemsep}{0pt} \settowidth
        {\labelwidth}{#1.}\sloppy}}{\end{list}}

\newcounter{itemlistc}
\newcounter{romanlistc}
\newcounter{alphlistc}
\newcounter{arabiclistc}

\newcommand{\fcaption}[1]{
        \refstepcounter{figure}
        \setbox\@tempboxa = \hbox{\footnotesize Fig.~\thefigure. #1}
        \ifdim \wd\@tempboxa > 5in
           {\begin{center}
        \parbox{5in}{\footnotesize\smalllineskip Fig.~\thefigure. #1}
            \end{center}}
        \else
             {\begin{center}
             {\footnotesize Fig.~\thefigure. #1}
              \end{center}}
        \fi}

\newcommand{\tcaption}[1]{
        \refstepcounter{table}
        \setbox\@tempboxa = \hbox{\footnotesize Table~\thetable. #1}
        \ifdim \wd\@tempboxa > 5in
           {\begin{center}
        \parbox{5in}{\footnotesize\smalllineskip Table~\thetable. #1}
            \end{center}}
        \else
             {\begin{center}
             {\footnotesize Table~\thetable. #1}
              \end{center}}
        \fi}

\def\@citex[#1]#2{\if@filesw\immediate\write\@auxout
        {\string\citation{#2}}\fi
\def\@citea{}\@cite{\@for\@citeb:=#2\do
        {\@citea\def\@citea{,}\@ifundefined
        {b@\@citeb}{{\bf ?}\@warning
        {Citation `\@citeb' on page \thepage \space undefined}}
        {\csname b@\@citeb\endcsname}}}{#1}}

\newif\if@cghi
\def\cite{\@cghitrue\@ifnextchar [{\@tempswatrue
        \@citex}{\@tempswafalse\@citex[]}}
\def\citelow{\@cghifalse\@ifnextchar [{\@tempswatrue
        \@citex}{\@tempswafalse\@citex[]}}
\def\@cite#1#2{{$\null^{#1}$\if@tempswa\typeout
        {IJCGA warning: optional citation argument
        ignored: `#2'} \fi}}

\def\pmb#1{\setbox0=\hbox{#1}
        \kern-.025em\copy0\kern-\wd0
        \kern.05em\copy0\kern-\wd0
        \kern-.025em\raise.0433em\box0}

\def\fnt#1#2{\footnotetext{\kern-.3em
        {$^{\mbox{\scriptsize #1}}$}{#2}}}

\def\fpage#1{\begingroup
\voffset=.3in
\thispagestyle{empty}\begin{table}[b]\centerline{\footnotesize #1}
        \end{table}\endgroup}

\def\runninghead#1#2{\pagestyle{myheadings}
\markboth{{\protect\footnotesize\it{\quad #1}}\hfill}
{\hfill{\protect\footnotesize\it{#2\quad}}}}
\font\tenrm=cmr10
\font\tenit=cmti10
\font\tenbf=cmbx10
\font\bfit=cmbxti10 at 10pt
\font\ninerm=cmr9
\font\nineit=cmti9
\font\ninebf=cmbx9
\font\eightrm=cmr8

\def\qed{\hbox{${\vcenter{\vbox{                        %HOLLOW SQUARE
   \hrule height 0.4pt\hbox{\vrule width 0.4pt height 6pt
   \kern5pt\vrule width 0.4pt}\hrule height 0.4pt}}}$}}

\def\bsc{{\sc a\kern-6.4pt\sc a\kern-6.4pt\sc a}}       %LATEX LOGO
\def\bflatex{\bf L\kern-.30em\raise.3ex\hbox{\bsc}\kern-.14em
T\kern-.1667em\lower.7ex\hbox{E}\kern-.125em X}

\def\beq{\begin{equation}}
\def\eeq{\end{equation}}
\def\p{\partial}

\def\la{\label}
\def\bea{\begin{eqnarray}}
\def\eea{\end{eqnarray}}
\begin{document}

\runninghead {P. Wiegmann}
{Bethe Ansatz and Classical Hirota Equation}

\normalsize\textlineskip
\thispagestyle{empty}
\setcounter{page}{1}

\copyrightheading{}                     %{Vol. 0, No. 0 (1993) 000---000}

\vspace*{0.88truein}

\fpage{1}
\centerline{\bf BETHE ANSATZ AND}
\vspace*{0.035truein}
\centerline{\bf CLASSICAL HIROTA EQUATION}
\vspace*{0.37truein}
\centerline{\footnotesize P.  WIEGMANN}
\vspace*{0.015truein}
\centerline{\footnotesize\it  James  Franck  Institute and  Enrico  Fermi
Institute}
\baselineskip=10pt
\centerline{\footnotesize\it University  of  Chicago,  Chicago, IL 60637,  USA}
%\baselineskip=10pt

%\vspace*{0.225truein}
%\publisher{(received date)}{(revised date)}

\vspace*{0.21truein}
\abstracts{We discuss an  interrelation between quantum integrable
models and classical
 soliton equations with discretized time. It appeared that  spectral
characteristics of quantum integrable systems may be obtained from
entirely classical set
up. Namely, the eigenvalues
of the quantum transfer matrix and the scattering $S$-matrix
itself are identified with a certain
$\tau$-functions of the discrete
Liouville equation.  The Bethe ansatz equations are
obtained as dynamics of zeros. For comparison
we also present the Bethe ansatz equations
 for elliptic solutions of the
  classical discrete Sine-Gordon equation.
   The paper is based on the recent study of classical
   integrable structures
    in quantum integrable systems.\cite{KLWZ}
 }{}{}

%\vspace*{10pt}
%\keywords{The contents of the keywords}

%\textlineskip                  %) USE THIS MEASUREMENT WHEN THERE IS
%\vspace*{12pt}                 %) NO SECTION HEADING

\vspace*{1pt}\textlineskip      %) USE THIS MEASUREMENT WHEN THERE IS
\section{Introduction  }
\vspace*{-0.5pt}
\noindent
1. In 1981 Hirota\cite{Hirota1} proposed a difference
equation which unifies
 all known continuous soliton equations.
A particular case of the
Hirota equation is a bilinear difference equation for a function $\tau
(n,l,m)$ of three
discrete variables:
\bea
&&\alpha \tau (n,l+1,m)\tau (n,l,m+1)+\beta \tau (n,l,m)\tau (n,l+1,m+1)
 \nonumber \\
&&\hskip 3.2cm
+\> \gamma \tau (n+1,l+1,m)\tau (n-1,l,m+1)=0\,,
\label{BDHE2}
\eea
where it is assumed that $\alpha +\beta +\gamma =0$.
Different continuum limits at different boundary conditions then reproduce
continuous
soliton equations (Kadomtsev-Petviashvili equation,
 Toda lattice, etc.). On the other hand,
$\tau (n,l,m)$ can be identified\cite{miwa} with the $\tau$-function
of a continuous hierarchy.

The
same equation (with a particular boundary condition) has suddenly appeared in
the theory of
{\it quantum} integrable systems as a
fusion relation for the {\it transfer matrix}
(trace of the quantum monodromy
matrix).\cite{KP,Kuniba1}

Thus classical integrability  emerges in quantum systems not
as a limiting procedure but rather as inherent and exact. It appears that
while solving an integrable quantum problem we in fact deal
with classical integrable equations, but with a {\it discretized time}.
Indeed, the Bethe ansatz equations and the eigenvalues of the S-matrix
and transfer matrix may be obtained by solving classical Hirota equation.

At present, we cannot treat an appearance of the classical
(discrete time) equations in quantum problem better than an
observation, although all this indicates that classical and
quantum integrable systems have much more in common than a
 regular classical limit $\hbar \rightarrow 0$.

2. The transfer matrix is one of the key objects in the theory of quantum
integrable systems.\cite{Faddeev}
Transfer matrices form a commutative family of operators
acting in the Hilbert space of a quantum problem. Let $R_{i,{\cal A}}(u)$ be
the $R$-matrix acting in the tensor product of
Hilbert spaces $V_i\otimes V_{\cal A}$.
Then the transfer matrix is a trace over the auxiliary space
$V_{\cal A}$ of the monodromy
matrix that is the matrix product of $N$\,$R$-matrices with the common
auxiliary space:
\bea {\hat T}_{\cal A}(u|y_i)&=
& R_{N,{\cal A}}(u-y_N)\cdots R_{2,{\cal A}}(u-y_2)
R_{1,{\cal A}}(u-y_1)\,, \nonumber\\
 T_{\cal A}(u)&=&
\mbox{tr}_{\cal A}{\hat T}_{\cal A}(u|y_i)\,.
 \la{transfer} \eea
 The transfer matrices
commute for all values of the spectral parameter $u$
 and different auxiliary spaces:
\beq
\phantom{A}[T_{\cal A}(u),\, T_{\cal A'}(u')]=0.
\label{F1}
\eeq
They can be diagonalized simultaneously, i.e. their eigenstates
do not depend on $\cal A$ and $u$. The family of eigenvalues of the
 transfer matrix is the first question to be addressed for an integrable
 system -
 the spectrum  of a quantum problem can be expressed in terms of
eigenvalues of the transfer matrix.

Moreover, the transfer matrix may be used to obtain the
scattering $S$-matrix. Indeed, at $u=0$ the fundamental
$R$-matrix becomes a permutation $R(u=0)=P$. Therefore,
choosing $\cal A$ to be the same as the representation of , say,
the first particle and setting $u=y_1$, the transfer matrix
 becomes the $S$-matrix of scattering the first particle
 with rapidity $y_1$ by the other $N-1$ particles
 \beq S_{\cal A}(u|y_2,...y_N)=
 R_{N,{\cal A}}(u-y_N)\ldots R_{2,{\cal A}}(u-y_2)
\,.
 \la{Smatrix} \eeq

3.  The transfer matrix corresponding to a given representation
in the auxiliary
space can be constructed out of transfer matrices
for some elementary space by
means of the {\it fusion procedure}.\cite{GL3,KS,JMO}
 The fusion procedure is
based on the fact that at certain values of the spectral parameter $u$
the $R$-matrix becomes essentially a
projector onto an irreducible representation space.
The fusion rules are especially simple in the $A_1$-case.
Consider for example the rational solution of the Yang-Baxter
equation for the fundamental (spinor)
representations of $SU(2)$:
\beq R_{1,1}(u)=u-2P\label{1000}\eeq
where $P$ is the permutation operator.
At a special value of the  spectral parameter  $u=\pm 2$,
the $R$-matrix becomes a projector onto irreducible moduli
of the tensor product  $[1/2]\otimes[1/2]= [0]\oplus[1]$ -
the singlet (spin-0)
at and onto the triplet (spin-1) :$R(\pm 2)=4P^{\mp}$. Then the transfer
matrix $T^1_2(u)$ with spin-1 auxiliary space is obtained
from the product of
two spin-1/2 monodromy matrices ${\hat T}^{1}_{1}(u)$
with arguments shifted by 2:
$$T^1_2(u)=
\mbox{tr}_{[1]}\big (R_{1,1}(-2){\hat T}^1_1(u+1){\hat
T}^1_1(u-1)R_{1,1}(-2)\big )\,.$$
A combination of the fusion procedure and the
Yang-Baxter equation results in
numerous functional relations (fusion rules) for the transfer matrix.\cite{GL3}
 They were recently combined into a universal bilinear form.\cite{KP,Kuniba1}
  The bilinear functional relations have
the most simple closed form for the models of the $A_{k-1}$-series and
representations corresponding to
{\it rectangular}
Young diagrams.

Let
$T^{a}_{s}(u)$ be the transfer matrix for the rectangular Young diagram
of length $a$ and height
$s$.
 They obey the following bilinear functional
 relation:
\beq
T^{a}_{s}(u+1)T^{a}_{s}(u-1)-
T^{a}_{s+1}(u)T^{a}_{s-1}(u)=
T^{a+1}_{s}(u)T^{a-1}_{s}(u)\,.
\label{F2}
\eeq
Since $T^{a}_{s}(u)$ commute at different $u,\,a,\,s$, the same equation
 holds for eigenvalues of the transfer matrices, so we can
 (and will) treat
$T^{a}_{s}(u)$ in   (\ref{F2}) as number-valued functions. The
bilinear fusion relations for models related to other
Dynkin graphs were suggested in ref.\cite{Kuniba1}.

Remarkably, the bilinear fusion relations (\ref{F2}) appear to be
identical to the
Hirota equation (\ref{BDHE2}). Indeed, one can eliminate the constants
$\alpha,\beta,\gamma$
by the transformation
$$
\tau (n,l,m)=\frac{ (-\alpha /\gamma)^{n^{2}/2}}
{(1+\gamma /\alpha )^{lm}}\> \tau _{n}(l,m),
$$
so that
\bea
&& \tau _n (l+1,m)\tau _n (l,m+1)- \tau _n (l,m)\tau _n (l+1,m+1) \nonumber \\
  &&\hskip 3.2cm - \>
 \tau _{n+1} (l+1,m)\tau _{n-1} (l,m+1)=0
\label{BDHE3}
\eea
 and then change variables from {\it light-cone} coordinates
$n,l,m$ to the {\it direct} variables
\bea
a=n,\;\;\;\;\;\; s=l+m,\;\;\;\;\;\; u=l-m-n,\nonumber\\
\tau _n (l,m)\equiv T^a_{l+m}(l-m-n).
\label{H1}
\eea
At least at a formal level, this transformation provides the
equivalence between (\ref{F2}), (\ref{BDHE2}) and (\ref{BDHE3}).

Relation between Hirota's equations and fusion relations goes much further.
It is known that a general class of Hirota's equations are related with the
geometry of Grassmannian manifolds and can be constructed out of a general
Young tableau.\cite{Sato,OHTI}
If, for example, the Young diagrams $Y=[a_1^{s_2},(a_1+a_2)^{s_1}]$
consist of two
rectangular blocks (one with $a_1$ lines of length $s_1 +s_2$ and the
second with
$a_2$ lines of length $s_1$), then the higher Hirota equations hold
\begin{eqnarray}
&&T_{s_1 , s_2 }^{a_1 , a_2 -1}(u)
T_{s_1 -1, s_2 -1}^{a_1 , a_2 +1}(u)+
T_{s_1 , s_2 +1}^{a_1 -1, a_2 -1}(u)
T_{s_1 , s_2 -1}^{a_1 +1, a_2 +1}(u) \nonumber \\
&+&T_{s_1 +1, s_2 }^{a_1 -1, a_2 }(u-1)
T_{s_1 -1, s_2 }^{a_1 +1, a_2 }(u+1) \nonumber \\
&=&T_{s_1 +1, s_2 }^{a_1 , a_2 -1}(u-1)
T_{s_1 -1, s_2 }^{a_1 , a_2 +1}(u+1)+
T_{s_1 , s_2 +1}^{a_1 -1, a_2 }(u-1)
T_{s_1 , s_2 -1}^{a_1 +1, a_2 }(u+1)\,.
\label{20}
\end{eqnarray}
This is also a fusion relation for the transfer matrix of the auxiliary
space of the
representation $Y$.

Leaving aside more fundamental aspects of this ``coincidence",
we exploit, as a first step, some technical advantages it offers.
Specifically, we
treat the functional relation (\ref{F2}) not as an identity but as a {\it
fundamental equation} which (together with particular boundary and
analytical conditions) completely determines all the
eigenvalues of the transfer matrix.  The
solution of the Hirota equation then appears in the form of the Bethe
ansatz
equations.
We anticipate that
this approach makes it possible to use some specific
tools of classical integrability and, in particular, the finite gap
integration technique.

The origin of $T^{a}_{s}(u)$ as an eigenvalue of the transfer matrix
(\ref{transfer})
imposes specific boundary
conditions:
\beq\label{bc0}T^a_s=0,\,\hskip 0.5cm a\ne 0,1,..,k\eeq
and, what is equally important, certain analytical properties. As a
general consequence of
the Yang-Baxter
equation,
the transfer matrices may be always normalized to be  {\it
elliptic
polynomials} in the spectral parameter, i.e. finite products of
Weierstrass
$\sigma$-functions (as in (\ref{Tgen}) below). Solutions with that
kind of analytical properties (called elliptic solutions) are well-known
in the theory of classical soliton equations due to the
works.\cite{AKM,Chud,kr1,bab}

4. To illustrate elliptic solutions in classical continuum
time soliton equations let us recall the result of the
Ref.\cite{kr1} for the  Kadomtsev-Petviashvili (KP) equation.
A particular elliptic solution can be written in the form
\beq
\label{KP}u(x,y,t)=2\sum_{i=1}^N{\cal P}(x-x_i(y,t))\eeq
where ${\cal P}$ is Weierstrass elliptic function and all
dynamics in $t$ and $y$ is hidden in the behaviour of
poles $x_i(y,t)$. A very existence of this ansatz is an
implementation of the integrability. It appears that poles
obey the elliptic Calogero-Moser many-body system with respect to $y$
\beq\label
{CM}
x_i^{\prime\prime}=-\sum_j {\cal P}^{\prime}(x_i-x_j).\eeq
Elliptic solutions can be lifted up to the discrete case.\cite{KLWZ,NRK}
 Then poles $x_i$
(or roots of the elliptic polynomial) obey a
fully discretized version of the Calogero-Moser model,
which can be recognized as the Bethe ansatz equations.

In this paper we shall consider the discrete version of
the classical Liouville equation. Its elliptic solutions
gives the eigenvalues of the transfer matrix and the
$S$-matrix of the quantum integrable systems associated with $A_1$ algebra.
For extension to $A_{k-1}$ algebra see  Ref.\cite{KLWZ}.
We also compare it, but do not study in detail, with Bethe ansatz
solution of the discrete Sine Gordon equation.\cite{KLWZ1}

The Hirota equation for the $A_1$ case may be obtained from the
 more general  (\ref{F2}), by setting the boundary conditions
\beq\label{bc1}T^a_s=0,\,\hskip 0.5cm a\ne 0,1,2\eeq
Then $T^{0,2}$ obey the discrete Laplace equation while
the  $T^1_s(u)\equiv T_s(u)$ obeys the discrete Liouville
equation (we continue to refer it as the Hirota equation):
\beq
T_s (u+1)T_s (u-1)-T_{s+1}(u)T_{s-1}(u)=T^{0}_{s}(u)T^{2}_{s}(u).
\label{T20}
\eeq
In fact the values of  $T^{0}_{s}(u)$ and $T^{2}_{s}(u)$
can be easily determined. They are the transfer matrices in
most symmetric and antisymmetric states of the $A_2$ case.
These states have dimension 1. In the most symmetrical
state there is no scattering, so $T^{0}_{s}(u)=1$. The
value of the transfer matrix in the most antisymmetric
state is known as the quantum determinant\cite{KRS}
\beq\label{det}T^{2}_{s}(u)=
 \frac{\gamma(u+s)\gamma(u+s+4)}{\gamma(u-s)\gamma(u-s+4)}.
\eeq
where
\beq \frac{\gamma(u+2)}{\gamma(u)}=\phi (u)\label{100}\eeq
and
\beq
\phi (u)=\prod _{k=1}^{N} \prod_{i=-q_k+1}^{q_k-1}\sigma (\eta (u-y_k+i ))\,.
\label{F8}
\eeq
Here $\sigma (x)$ is the Weierstrass $\sigma$-function
and $q_k$ is a spin of the $k$th quantum space. The
roots $y_k$ and degree $N$ of the elliptic polynomial
$\phi (u)$ and parameters of the
Weierstrass function are parameters of the quantum model.
Below we concentrate mostly on the simplest case $q_k=1$.

Apparently, different solutions of   (\ref{T20})  have a
number common zeros and $T_s(u)$ at general parameters $y_k$
are polynomials of  degree $s\sum q_k$. These zeros
 can be gauged out, so all $T_s(u)$ (different
eigenvalues at different $s$ and $q_k$) remain
polynomials of the same degree $N $. The maximal gauge transformation is
$$T_s(u)\rightarrow T_s(u)\prod_{p=2}^s\phi(u+s-2p)
\prod_{k=1}^N\prod_{p=1}^{q_k-s}\sigma(\eta (u-s+q_k+2-2p)$$
where the last product holds for $q_k>s$.
This gauge transformation simplifies the r.h.s. of  (\ref{T20})
\beq
T_s (u+1)T_s (u-1)-T_{s+1}(u)T_{s-1}(u)=\phi (u+s) \phi (u-s-2).
\label{T2}
\eeq
In this normalization all $T_s$ are polynomials of
the same degree with no common zeros.

A general solution of the discrete Liouville equation (\ref{T20}) is
parametrized by two arbitrary functions of one variable.
Not all of them correspond to
eigenvalues of the quantum transfer matrix, but only
elliptic solution. Elliptic polynomials may be characterized
by  its roots $z_{j}^{(s)}$
\beq
T_{s}(u)=A_{s}e^{\mu_s u}
\prod _{j=1}^{N} \sigma (\eta (u-z_{j}^{(s)}))\,,
\label{Tgen}
\eeq
where  $A_{s}$, $\mu_s$ do not depend on $u$. Similar to
 the continuum case, roots obey some dynamics in a discrete
 ``time" variable $s$. It is determined by the Bethe-ansatz-like
 equations
\beq
\frac{\phi (z_{j}^{(s)}+s+1)}
{\phi (z_{j}^{(s)}+s+3)}=-\frac{A_{s-1}}{A_{s+1}}
\prod _{k=1}^{N}
\frac{\sigma(\eta(z_{j}^{(s)}-z_{k}^{(s-1)}-1))}
{\sigma(\eta(z_{j}^{(s)}-z_{k}^{(s+1)}+1))}\,,
\label{R11a}
\eeq
\beq
\frac{\phi (z_{j}^{(s)}-s+1)}
{\phi (z_{j}^{(s)}-s-1)}=-\frac{A_{s-1}}{A_{s+1}}
\prod _{k=1}^{N}
\frac{\sigma(\eta(z_{j}^{(s)}-z_{k}^{(s-1)}+1))}
{\sigma(\eta(z_{j}^{(s)}-z_{k}^{(s+1)}-1))}\,.
\label{R11b}
\eeq
The constant $A_s$ can be easily found in the rational and trigonometric
case by comparing
the leading powers of polynomials in the (\ref{T2}). In the rational case
it obeys the equation
$$A_s^2-A_{s+1}A_{s-1}=0.$$
 Each solution
of these equations gives an eigenvalue of
 the transfer matrix. Equations (\ref{R11a}) and (\ref{R11b})
 determine the zeros of the transfer matrix.  The
 traditional Bethe ansatz equations
\beq
e^{-4\eta \nu}\frac{\phi (u_j )}{\phi (u_j -2)}=-\prod _{k}
\frac{\sigma(\eta (u_j-u_k +2))}{\sigma(\eta(u_j-u_k-2))}\,.
\label{Bethe0}\eeq
corresponding to
 zeros of the Baxter $Q$ matrix are discussed in  Sec. 5.

\section{Other Forms of the Hirota Equation}
\noindent
The equation (\ref{T20}) is known as a discrete
version of the Liouville equation\cite{Hirota2}
 written in terms of the $\tau$-function.
It can be recast to a  more universal form in terms of the
discrete Liouville
field
\beq
Y_{s}^{1}(u)\equiv
Y_s (u)=\frac{ T_{s+1}(u)T_{s-1}(u)}{\phi (u+s)\phi (u-s-2)}
\label{Y1}
\eeq
which hides the function $\phi(u)$ in the r.h.s. of (\ref{T2}).
The equation becomes
\beq
Y_s (u-1)Y_s (u+1)=(Y_{s+1}(u)+1)(Y_{s-1}(u)+1)\,.
\label{Ysys}
\eeq
with a boundary condition $Y_0(u)=0$.
(Let us note that the same functional equation but with different
analytic properties of the solutions
appears in the thermodynamic Bethe ansatz.\cite{Zam})

The functional equations (\ref{Ysys}) can be further
rewritten in an integral form. Introducing
$\varepsilon_s(u)=\ln Y_s(u)$, one gets
\beq\varepsilon_s(u)-\int_{-i\infty}^{i\infty}\frac{1}{2\pi
\cos \frac{\pi}{2}(u-u^\prime)}
\log((1+e^\varepsilon_{s+1})(1+e^\varepsilon_{s-1}))=
\sum_k
\delta_{s q_k}\bar\varphi_k(u)\label{1}\eeq
where $\bar\varphi_k(u)$ are some known functions
computed out of the r.h.s. of the  (\ref{T20}),
such as $\bar\varphi_k(u+1)+\bar\varphi_k(u-1)=0$.\cite{KP}

\newpage
\section{Difference and Continuum $A_1$-Equation}
\noindent
{\it Liouville equation.}
In the continuum limit one should put $Y_s (u)=\delta ^{-2}
%\exp (-\varphi (x,t))$,
e^{-\varphi (x,t)}$,
$u=\delta^{-1}x,\,s=\delta^{-1}t$. An expansion in
$\delta\rightarrow 0$ then
gives the continuous Liouville equation
\beq
\p_{s}^{2}\varphi-\p_{u}^{2}\varphi =2\exp (\varphi )\,.
\label{Liouv}
\eeq
{\it Sine-Gordon equation.} For illustrative purposes let us compare it with
discrete Sine-Gordon (SG) equation, which
requires quasi-periodic  boundary condition with respect to the
Dynkin diagram\cite{Hirota3}
\beq
T_{s}^{a+1}(u)=e^\alpha \lambda^{2a}T_{s}^{a-1}(u-2),
\label{sg2}
\eeq
where $\alpha$ and $\lambda$ are parameters. Plugging this condition into
(\ref{F2}), we get:
\beq
T_{s}^{1}(u+1)T_{s}^{1}(u-1)-T_{s+1}^{1}(u)T_{s-1}^{1}(u)=
e^\alpha \lambda^2 T_{s}^{0}(u)T_{s}^{0}(u-2),
\label{sg3a}
\eeq
\beq
T_{s}^{0}(u+1)T_{s}^{0}(u-1)-T_{s+1}^{0}(u)T_{s-1}^{0}(u)=
e^{-\alpha} T_{s}^{1}(u)T_{s}^{1}(u+2).
\label{sg3b}
\eeq
Let us introduce two fields $\rho ^{s,u}$ and $\varphi ^{s,u}$ on the
square $(s,u)$ lattice
\beq
T_{s}^{0}(u)=\exp(\rho ^{s,u}+\varphi ^{s,u}),
\label{sg4a}
\eeq
\beq
T_{s}^{1}(u+1)=\lambda^{1/2}\exp(\rho ^{s,u}-\varphi ^{s,u}),
\label{sg4b}
\eeq
and substitute them into (\ref{sg3a}) and (\ref{sg3b}). Finally, eliminating
$\rho ^{s,u}$, one gets the discrete SG equation:
\bea
&&\sinh\> (\varphi ^{s+1,u}+\varphi ^{s-1, u}
-\varphi ^{s,u+1}-\varphi ^{s, u-1}) \nonumber \\
&&\hskip 2cm =\lambda
\sinh\> (\varphi ^{s+1,u}+\varphi ^{s-1, u}
+\varphi ^{s,u+1}+\varphi ^{s, u-1}+\alpha)\,.
\label{sg5}
\eea
The constant $\alpha$ can be removed by the redefinition
$\varphi ^{s,u}\rightarrow \varphi ^{s,u}-\alpha/4$.

Another useful form of the discrete SG equation appears in variables
\beq\label{X1}
X^a_s(u)=-
\frac{T_{s}^{a}(u+1)T_{s}^{a}(u-1)}
{T_{s}^{a+1}(u)T_{s}^{a-1}(u)}=-1-Y_{s}^{a}(u)
\label{X}
\eeq
Under condition (\ref{sg2}) one has
\beq \la{X2}X^{a+1}_s(u)=X^{a-1}_{s}(u-2),\;\;\;
\;\; \lambda ^{2}X^{a+1}_{s}(u+1)X^{a}_{s}(u)=1\,,
\eeq
so there is only one independent function
\beq
X_{s}^{1}(u)\equiv x_{s}(u)=-e^{-\alpha}\lambda ^{-1}
\exp \big (-2\varphi ^{s,u}-2\varphi ^{s,u-2}\big )\,.
\label{x}
\eeq
The discrete SG equation becomes\cite{Hirota3,FV,pendulum}
\beq\la{sg6}
x_{s+1}(u)x_{s-1}(u)=\frac
{(\lambda+
x_{s}(u+1))(\lambda +x_{s}(u-1))}
{(1+\lambda x_{s}(u+1))(1+\lambda
x_{s}(u-1))}\,.
\eeq
In the limit $\lambda\rightarrow 0$   (\ref{sg6}) turns into the
discrete Liouville equation (\ref{Ysys}) for $Y_s (u)=
-1-\lambda ^{-1}x_{s}(u)$.

\section{Determinant representations}
\noindent
Relation between the Hirota equation and Plucker relations of the
coordinates of the
Grassmannian manifolds suggests a numerous determinant representations of its
solutions.
The most familiar one\cite{BR2}  allows one to
express $T_{s}^{a}(u)$ through $T_{1}^{a}(u)$ or $T_{s}^{1}(u)$. For
instance the determinant
formula  giving the evolution in $a$ holds
\beq
T_{s}^{a}(u)=\det \big (T_{1}^{a+i-j}(u+i+j-s-1)\big ),
\;\;\;\;\; i,j=1,\ldots ,s\,, \;\;\;\;\; T^{a}_{0}(u)=1\,.
\label{Tdet1}
\eeq

Another representation appeared to be more suitable for particularities of
the $A_{k-1}$
Liouville equation (\ref{F2}) and (\ref{bc0}). They explicitly express its
solution in terms of
$2k$ arbitrary functions of one variable\cite{KLWZ} as a determinant of
the $k\times k$
matrix
\bea \la{tdet1}
\tau_a (l,m)&=&\det M_{ij}\,, \nonumber\\
M_{ji}&=&\left\{\begin{array}{ll}
h_i(u+s+a+2j)&\mbox{if $j=1,...,k-a;\;i=1,...,k$}\\
\bar h_i(u-s+a+2j)&\mbox{if $j=k-a+1,...,k;\; i=1,..., k$}
\end{array}\right. \eea

In the $A_1$ case the determinant has a particular simple form:
\beq
T^0_s(u)\equiv\phi (u+s)=\left |
\begin{array}{ll}R(u+s) & Q(u+s)\\
R(u+2+s) & Q(u+2+s) \end{array} \right |,
\label{B1.1}
\eeq
\beq
T^2_s(u)\equiv\phi (u-s-2)=\left |
\begin{array}{ll}\bar R(u-s-1) & \bar Q(u-s-1)\\
\bar R(u-s+1) & \bar Q(u-s+1) \end{array} \right |.
\label{B1.2}
\eeq
\beq
T_{s}(u)=
\left |
\begin{array}{ll}Q(u+s+1) & R(u+s+1)\\
\bar Q(u-s) & \bar R(u-s) \end{array} \right |.
\label{B1.5b}
\eeq
Compare  first two equations one finds
\beq
\bar Q(u)=Q(u-1),\;\;\;\;\;\; \bar R(u)=R(u-1)\,,
\label{QR}
\eeq
so that
\beq
T_s (u)=
\left | \begin{array}{ll}Q(u+s+1)& R(u+s+1)\\
Q(u-s-1)& R(u-s-1) \end{array}\right |\,.
\label{A1.5c}
\eeq

\section{Bethe-Ansatz}
\noindent
The functions $Q,R$  are further determined by the r.h.s. of the
(\ref{T2}) and by the analyticity requirement that $T_s(u)$ is an
elliptic polynomial
(\ref{Tgen}). One can show that it  also means that $Q(u)$ or  $R(u)$ (but
not necessarily
all together) is an elliptic polynomials of degree $N/2$ multiplied by
exponential function
\beq\label{Q} Q(u)=e^{\nu\eta u}\prod_{j=1}^{M}\sigma(\eta(u-u_j))\eeq

Let $u_j$,  $j=1,\ldots ,M$ be
zeros of $Q(u)$. Then, evaluating (\ref{B1.1})
at
$u=u_j$,
$u=u_j -2$, we obtain the relations
\beq
\phi (u_j )=Q(u_j +2)R(u_j )\,, \;\;\;\;\;\;\;
\phi (u_j -2)=-Q(u_j -2)R(u_j )\,,
\label{QRQR}
\eeq
whence it holds
\beq
\frac{\phi (u_j )}{\phi (u_j -2)}=-
\frac{Q(u_j +2)}{Q(u_j -2)}\,,
\label{Bethe1}
\eeq
Equation (\ref{Bethe1}) is the celebrated Bethe ansatz equation (\ref{Bethe0}).

In the elliptic case the degrees of the elliptic polynomial $Q(u)$  (for
even $N$) is equal
to $M=N/2$, provided that $\eta$ is incommensurable with the lattice
spanned by $\omega _1$,
$\omega _2$. In the trigonometric and rational cases there are no such
strong restrictions
on degrees $M$ and $\tilde M$ of $Q$. This is
because a part of their zeros may tend to infinity thus reducing the
degree. Whence $M$ and $\tilde M$ can be arbitrary integers not
exceeding $N$. However, they must be complement to each other:
$M+\tilde M =N$. The traditional choice is $M\leq N/2$. In particular,
the solution $Q(u)=1$ ($M=0$) corresponds to the
simplest reference state
(``bare vacuum") of the model.

\section{Linear Problem}
\noindent
The function $Q$  naturally appears as the object of the linear problem of the
discrete Liouville equation.
Indeed, the equation (\ref{T20}) admits a zero
curvature representation. Let us
start from a more general equation
\beq
T_s (u+1)T_s (u-1)-T_{s+1}(u)T_{s-1}(u)=\phi (u+s)\bar \phi (u-s),
\label{T0}
\eeq
with two independent functions $\phi$, $\bar \phi$ and later
impose the relation
$\bar \phi (u)=\phi(u-2)$.
The nonlinear equation appears as a consistency conditions of the
following auxiliary
linear problems
\beq
T_{s+1}(u) Q(u+s)-T_{s}(u-1)Q(u+s+2)=\phi (u+s) \bar Q(u-s-1),
\label{T5a}
\eeq
\beq
T_{s+1}(u) \bar Q(u-s+1)-T_{s}(u+1) \bar Q(u-s-1)=\bar\phi (u-s) Q(u+s+2)\,,
\label{T5b}
\eeq
where we introduced two functions of one variable $Q(u)$ and
$\bar Q(u)$.  These are the functions appeared in the determinant
representations
(\ref{B1.1}), (\ref{B1.2}), and (\ref{B1.5b}).

The fact that
functions of the linear problem depend on one variable
is a specific feature of the Liouville
equation.

To compare let us present the linear problem for the
discrete Sine-Gordon equation (\ref{sg3a}) and ({\ref{sg3b}). It reads
 \beq
T^1_{s+1}(u) F^0_s(u)-T^1_{s}(u-1)F^0_{s+1}(u+1)=T^0_s(u)F^1_{s+1}(u),
\label{sg7a}
\eeq
\beq
T^0_{s+1}(u-2) F^1_s(u)-T^0_{s}(u-3)F^1_{s+1}(u+1)=T^1_s(u)F^0_{s+1}(u-2)
\label{sg7b}
\eeq
 \beq
T^0_{s+1}(u) F^0_s(u)-T^0_{s}(u-1)F^0_{s+1}(u-1)
=e^{-\alpha}T^1_s(u)F^1_{s+1}(u+2),
\label{sg8a}
\eeq
\beq
T^1_{s+1}(u-1) F^1_s(u)-T^1_{s}(u)F^1_{s+1}(u-1)
=e^{\alpha}\lambda^2 T^0_s(u-3)F^0_{s+1}(u)
\label{sg8b}
\eeq
In this case the wave functions $F^0_s(u)$ and
$F^1_s(u)$ depend on both $s$ and $u$.

\section{The Baxter Relation}
\noindent
A general solution of the discrete Liouville equation (for arbitrary $\phi$
and $\bar\phi$) may be expressed through two independent functions $Q(u)$
and $\bar Q(u)$.
Below we follow the same lines developed for solving
continuous classical Liouville equation  (see, e.g., Refs.\cite{Gervais,Pogreb}
and references therein).
First we rearrange   (\ref{T5a}) and (\ref{T5b}) as
\beq
\phi (u-2)Q(u+2)+\phi (u)Q(u-2)=A(u)Q(u),
\label{T13}
\eeq
\beq
\bar \phi (u)\bar Q(u+3)+\bar \phi (u+2)\bar Q(u-1)=\bar A(u)\bar Q(u+1),
\label{T14}
\eeq
where we have introduced the quantities
\beq
A(u)=\frac{\phi (u-2)T_{s+1}(u-s)+\phi (u)T_{s-1}(u-s-2)}
{T_s (u-s-1)}\,,
\label{T16}
\eeq
\beq
{\bar A}(u)=\frac{\bar \phi (u+2)T_{s+1}(u+s)+\bar \phi (u)T_{s-1}(u+s+2)}
{T_s (u+s+1)}\,.
\label{T17}
\eeq
Due to the consistency condition
(\ref{T0}),  $A(u)$ and $\bar A(u)$ are functions of one variable and do
not depend on
$s$. The symmetry between $u$ and $s$ allows one to construct similar
objects which in
their turn do not depend on $u$.
Functions $A(u)$ and $\bar A(u)$, in the r.h.s. of (\ref{T13}) and
(\ref{T14}) are
the conservation
laws of the $s$-dynamics.

Running ahead, let us note that the connection between $\phi$ and
$\bar\phi$, $\bar \phi (u)=\phi (u-2)$,
and its consequence  $T_{-1}(u)=0$, simplifies
(\ref{T13}) - (\ref{T17}).
Putting $s=0$ and using the  boundary condition $T_{-1}(u)=0$, we find
\beq
A(u)=\bar A(u)=T_1 (u)\,.
\label{T8}
\eeq
Therefore, we have the relations
\beq
T_s (u-1)T_1 (u+s)=\phi (u+s-2)T_{s+1}(u)+\phi (u+s)T_{s-1}(u-2),
\label{fus1}
\eeq
\beq
T_s (u+1)T_1 (u-s)=\phi (u-s)T_{s+1}(u)+\phi (u-s-2)T_{s-1}(u+2),
\label{fus2}
\eeq
\beq
\phi (u-2)Q(u+2)+\phi (u)Q(u-2)=T_1 (u)Q(u)\,.
\label{T6}
\eeq
These equalities are also known as fusion relations and
have been obtained in Refs.\cite{GL3,KR,BR1}, and    (\ref{T6}) is also
known as the
Baxter $T$-$Q$ relation.\cite{Baxter}
So the Baxter $Q$ function and  $T$-$Q$ relation naturally appear in the
context of the auxiliary
linear problems for the bilinear fusion relation.

The Baxter equation (\ref{T6}) gives rise to
a generalized spectral problem for a  discrete
operator of the second order: {\it find all elliptic
``eigenvalues"of the operator $\phi (u-2)e^{2\partial_u
}+\phi (u))e^{-2\partial_u}$}. One may show that in
 this case the eigenfunction $Q(u)$ itself is an elliptic polynomial.
%Particular case of this problem when $T_1(u)=const$ has been considered in
%\cite{WZ}.
The Bethe ansatz is a practical tool to solve this problem.

Let us consider   (\ref{T13}) (resp. (\ref{T14})) as a second order
linear difference equation,
 where the function
$A(u)$ ($\bar A(u)$) is determined from the initial data. Let $R(u)$
(resp.
$\bar R (u)$) be a second (linearly independent) solution of
(\ref{T13}) (resp.
(\ref{T14})) normalized so that the wronskians are
\beq
W(u)=\left |
\begin{array}{ll}R(u) & Q(u)\\

R(u+2) & Q(u+2) \end{array} \right |=\phi (u),
\label{A1.1}
\eeq
\beq
\bar W(u)=\left |
\begin{array}{ll}\bar R(u) & \bar Q(u)\\
\bar R(u+2) & \bar Q(u+2) \end{array} \right |=\bar\phi (u+1).
\label{A1.2}
\eeq
Then the general solution of (\ref{T0}) is given by the determinant
(\ref{B1.5b}).
%\beq
%T_{s}(u)=
%\left |
%\begin{array}{ll}Q(u+s+1) & R(u+s+1)\\
%\bar Q(u-s) & \bar R(u-s) \end{array} \right |,
%\label{A1.5b}
%\eeq

For any given $Q(u)$ and $\bar Q(u)$ the
second solution $R(u)$ and $\bar R(u)$
(defined modulo a linear transformation $R(u)\rightarrow R(u)+\alpha
Q(u)\,$) can be
explicitly found out of the first order recurrence relations
(\ref{B1.1}),
(\ref{B1.2}), if necessary. Then (\ref{B1.5b}) determines $T_s(u)$ only
in terms of
$Q(u)$:
\beq
T_s (u)=Q(u+s+1)Q(u-s-1)
\sum _{j=0}^{s}\frac {\phi (u-s+2j-1)}{Q(u-s+2j+1)Q(u-s+2j-1)}\,.
\label{T12}
\eeq
This formula has been obtained in Refs.\cite{KR,BR1} by direct resolving
fusion recurrence relations (\ref{fus1}) and (\ref{fus2}).
 The Bethe ansatz equations may be also obtained as a conditions of
cancelation of poles
in (\ref{T12}).

Let us list some more useful
representations.

The conserved quantities $A(u)$ and $\bar A(u)$ in terms of $Q$ and $R$
have the form:
\beq A(u)=Q(u+2)R(u-2)-R(u+2)Q(u-2), \label{rel1}
\eeq
\beq
\bar A(u)=\bar R(u+3)\bar Q(u-1)-\bar Q(u+3)\bar R(u-1),
\label{rel2}
\eeq
which are direct corollaries of (\ref{T13}) and (\ref{T14}).

%\section{Equivalent Forms of the Baxter Equation}

In its turn Baxter's relation (\ref{T6})
and its ``chiral" versions (\ref{T13}) and (\ref{T14}) also enjoy some
determinant
representation:\cite{KLWZ}
\beq
\left | \begin{array}{lll} T_s (u)& T_{s+1}(u-1)& Q(u+s+1)\\ &&\\
T_{s+1}(u+1)& T_{s+2}(u)& Q(u+s+3)\\  &&\\
T_{s+2}(u+2)& T_{s+3}(u+1)& Q(u+s+5) \end{array} \right |=0\,,
\label{B1a}
\eeq
\beq
\left | \begin{array}{lll} T_s (u)& T_{s+1}(u+1)& \bar Q(u-s)\\ &&\\
T_{s+1}(u-1)& T_{s+2}(u)& \bar Q(u-s-2)\\ &&\\
T_{s+2}(u-2)& T_{s+3}(u-1)& \bar Q(u-s-4) \end{array} \right |=0\,,
\label{B1b}
\eeq
provided that $T_s(u)$ obeys the Hirota equation.

\newpage
\section{Zeros of the transfer matrix}
\noindent
The linear problem  (\ref{T5a}) and (\ref{T5b}) are convenient to find  the
Bethe-ansatz-like
equations (\ref{Tgen}) for
zeros of $T_s(u)$.  Substituting $u=z_i^{(s+1)}$ and $u=z_i^{(s)}+1$ into
(\ref{T5a}) and
$u=z_i^{(s+1)}$ and $u=z_i^{(s)}-1$ into (\ref{T5b}) we obtained the
(\ref{R11a}) and
(\ref{R11b}) referred in the introduction. In the rational case
($\sigma(\eta u)\rightarrow u$) $A_s =s+1$.

Let us compare the Bethe-ansatz-like equations  of the Liouville
equation (\ref{R11a}) and (\ref{R11b}) with the Sine-Gordon equation
(\ref{sg3a}) and (\ref{sg3b}). Let $z^0_i(s)$ and $z^1_j(s)$
be the zeros of the elliptic polynomial
\beq
T^{1,2}_{s}(u)=A^{1,2}_{s}e^{\mu (s)u}
\prod _{j=1}^{N} \sigma (\eta (u-z^{1,2}_{j}(s)))\,.
\label{Tgen1}
\eeq
Substituting $u=z^{1,2}_i(s)$ into the second pair of equations
of the linear problem (\ref{sg8a}) and (\ref{sg8b}) and excluding $F$'s
after some algebra one obtains
\bea
&&\prod _{j=1}^{N}
\biggl(\frac{\sigma(\eta(z_{i}^{1}(s+1)-z_{j}^0(s+1)-1)}
{\sigma(\eta(z_{i}^1(s+1)-z_{j}^0(s+2)-3)} \biggr)
\biggl( \frac{\sigma(\eta(z_{i}^{1}(s)-z_{j}^1(s)+1)}
{\sigma(\eta(z_{i}^1(s+1)-z_{j}^1(s+1)+1)} \biggr) \nonumber \\
&& \hskip 3cm \times \frac{\sigma(\eta(z_{i}^{1}(s+1)-z_{j}^0(s+2)-2)}
{\sigma(\eta(z_{i}^1(s+1)-z_{j}^0(s))}
=-\frac{A^0_sA^1_{s+1}}{A^0_{s+1}A^1_{s}}
\label{R12a}
\eea
\bea
&&\prod _{j=1}^{N}
\biggl(\frac{\sigma(\eta(z_{i}^{0}(s+1)-z_{j}^1(s+1)+2)}
{\sigma(\eta(z_{i}^0(s+1)-z_{j}^1(s+1)+1)}\biggr)
\biggl(\frac{\sigma(\eta(z_{i}^{0}(s)-z_{j}^0(s+1)+1)}
{\sigma(\eta(z_{i}^0(s)-z_{j}^0(s)+1)}\biggr) \nonumber  \\
&& \hskip 3cm \times \frac{\sigma(\eta(z_{i}^{0}(s+1)-z_{j}^1(s+2)-1)}
{\sigma(\eta(z_{i}^0(s+2)-z_{j}^1(s+2))}
=-\frac{A^0_sA^1_{s+1}}{A^0_{s+1}A^1_{s}}
\label{R12b}
\eea
The constants $A_s$ may be easily found in the rational or trigonometric
limits by
comparing leading orders of polynomials in the Hirota equation. In the
rational case one
gets:
\beq\label{13}
(A^1_s)^2-A^1_{s+1}A^1_{s-1}=e^\alpha\lambda^2(A^0_s)^2
\eeq
\beq\label{15}
(A^0_s)^2-A^0_{s+1}A^0_{s-1}=e^{-\alpha}\lambda^2(A^1_s)^2.\eeq
These equations gives a discrete dynamics of zeros in $s$.
They are to be compared with
dynamics of zeros of elliptic solutions of classical nonlinear
equations.\cite{AKM,Chud,kr1,bab}

\section{Crossing Symmetry and the S-Matrix}
\noindent
As it was noticed in the introduction, once the eigenvalues
of the transfer matrix are found it is easy to get eigenvalues
of the  $S$-matrix (\ref{Smatrix}). In relativistic field theories
the  $S$-matrix is normalized in  a way to be unitary and
crossinvariant, i.e. to be unitary in both $s$ and $t$ channels.

Let us consider for instance the spin 1/2 fundamental
$A_1$ rational  $S$-matrix.
It has the form:
\beq\la{111}S(u)=U(u)R(u)\eeq
where the $R$-matrix is given by    (\ref{1000}) and the amplitude
  $U(u)$ is  restricted by the unitarity and the cross symmetry.
  The unitarity conditions in both channels are\cite{ZZ}
\beq\label{22}U(u)U^*(-u)=\frac{1}{\phi(u+2)\phi(-u+2)}\eeq
\beq\label{33}\frac{U(u+2)}{U(u-2)}=\frac{\phi(u+2)\phi(u-2)}
{\phi(u)\phi(u+4)}.\eeq
  The minimal solution is
\beq\la{7}U(u)=\frac{\Gamma(\frac{u+4}{4})\Gamma(\frac{2-u}{4})}
{(u+2)\Gamma(\frac{4-u}{4})\Gamma(\frac{u+2}{4})}.\eeq

The Hirota equation (\ref{T2}) has a natural normalization which
gives the cross-invariant unitary  $S$-matrices in all representations.
 It is a gauge which brings the quantum determinant (\ref{det})
  to 1 and the Hirota equation and the Baxter relation (\ref{T6})
  to the universal form
\beq
\tau_s (u+1)\tau_s (u-1)-\tau_{s+1}(u)\tau_{s-1}(u)=1
\label{t2}
\eeq
 \beq
\tilde Q(u+2)+\tilde  Q(u-2)=\tau_1 (u)\tilde Q(u)\,.
\label{14}
\eeq
 This is achieved by the gauge
\beq\la{11}
T_s(u)=\tau_s(u)\frac{f(u+s)}{f(u-s)}\eeq
where
\beq\la{12}f(u+1)f(u-1)=\gamma (u)\gamma(u+4).\eeq
Then $\tau_s$ is a meromorphic function with prescribed positions of poles.
It brings the $R$-matrix to the cross unitary $S$-matrix
$$S(u)=\frac{f(u+1)}{f(u-1)}R(u).$$
This factor can be identified with the  factor $U$ in  (\ref{1}). Indeed,
  (\ref{12}) implies
$$\frac{f(u+3)}{f(u-1)}=\phi(u)\phi(u+4)$$
$$\frac{f(u+3)f(u-3)}{f(u+1)f(u-1)}=\frac{\phi(u)\phi(u+4)}{\phi(u-2)
\phi(u+2)}$$
This is equivalent to the unitarity and cross-unitarity relation
(\ref{22}) and (\ref{33}).

\section{Conclusion and Outlook}
\noindent
It turned out that  quantum integrable
models have a deep
connection with classical discrete soliton equations:
 the fusion rules for quantum
transfer matrices are identical to the Hirota  bilinear
difference equation with a certain boundary conditions
and analytical requirement. This coincidence goes far
beyond the simplest example considered in this paper.
It most likely holds  for all Lie algebras and representations
and all integrable boundary conditions.

Under this identification  eigenvalues of the transfer matrix
are represented as $\tau$-functions. Positions of zeroes of
the solution is determined by the Bethe ansatz
equations.  We therefore conclude that the information one
 usually obtains from the Yang-Baxter equation may be also
 found from classical discrete time equations, without
  implying a quantization.

In general the Bethe ansatz technique, what has been thought
of as a specific tool of
quantum integrability  is shown to
exist in classical discrete nonlinear integrable equations.
 We demonstrated this on two examples - discrete Liouville and
 discrete Sine-Gordon equation.  It is tempting to compare
solutions of the quantum Sine-Gordon field theory and discrete
classical Sine-Gordon equation.
In a forthcoming paper we extend the theory of elliptic
solution known in continuum soliton equations to a more
general class of discrete equations by means of the Bethe ansatz.

The difference bilinear equation, although with different
analytical requirements, has
 appeared in quantum integrable systems in another context.
 Spin-spin correlation
functions of the Ising model obey a bilinear
difference equation that can be recast into the form of
Hirota equation.\cite{CoyWu,Perk1,Perk2}
More recently, nonlinear equations
for correlation functions have been derived for more general class of
 quantum integrable models, by virtue a new approach of the Ref.\cite{book}.
Thermodynamic Bethe ansatz  equations written in a
 form of functional relations\cite{Zam} (see, e.g., Ref.\cite{BLZ}) appeared
to be identical with Hirota equation (although with another
analytic properties) as well.

 All these suggest that the Hirota equation may play a role of
 a master equation simultaneously for
both classical and quantum integrable systems, such as the
``equivalence" between quantum
systems and discrete classical dynamics may be extended beyond
the spectral properties
discussed in this paper.

\nonumsection{Acknowledgements}
\noindent
This work has been supported in part by NSF grant DMR-9509533. The author
thanks A. Luther
for his hospitality in NORDITA in August 1996 where this paper has been
written.
%\newpage

\nonumsection{References}

\end{document}